\documentclass[preprint]{iucr}  % DO NOT 
\RequirePackage{graphicx} %DELETE THIS LINE

     \journalcode{J}              % Indicate the journal to which submitted
\usepackage{amsmath,amssymb}
\usepackage{xcolor}

\begin{document}  % DO NOT DELETE THIS LINE

     %-------------------------------------------------------------------------
     % The introductory (header) part of the paper
     %-------------------------------------------------------------------------

     % The title of the paper. Use \shorttitle to indicate an abbreviated title
     % for use in running heads (you will need to uncomment it).

\title{Neutron Scattering Cross-Section Correction Incorporating Neutron Wavelength Effects}

     % Authors' names and addresses. Use \cauthor for the main (contact) author.
     % Use \author for all other authors. Use \aff for authors' affiliations.
     % Use lower-case letters in square brackets to link authors to their
     % affiliations; if there is only one affiliation address, remove the [a].

\author[a]{Karrie E.}{An}
\author[b]{Guan-Rong}{Huang}
\author[c]{Changwoo}{Do}
\author[c]{Wei-Ren}{Chen}

\aff[a] {Neutron Technologies Division, Oak Ridge National Laboratory, Oak Ridge, TN 37831, \country{United States}}
\aff[b]{Department of Materials and Optoelectronic Science, National Sun Yat-sen University, Kaohsiung 80424, \country{Taiwan}}
\aff[c]{Neutron Scattering Division, Oak Ridge National Laboratory, Oak Ridge, TN 37831, \country{United States}}

%\shortauthor{Soape, Author and Doe}

\maketitle   % DO NOT DELETE THIS LINE

\begin{synopsis}
In this research, we present a methodology for addressing the numerical accuracy of neutron scattering cross-sections for commonly encountered elements found in soft materials. Our approach involves taking into consideration the contributions weighted by wavelength across a spectrum of low neutron energy levels commonly used in typical neutron scattering experiments.
\end{synopsis}

\clearpage

\begin{abstract}
This study outlines a numerical methodology aimed at rectifying the neutron scattering cross-sections of fundamental elements across a range of low neutron energies typically employed in general neutron scattering experiments. By using the experimental power law relationship governing the cross-section's dependence on neutron wavelength, we establish a mathematical connection between these two variables. Leveraging this relationship, the scheme of central moment expansion is adopted to correct the cross-sections that are applicable to general neutron wavelength distributions commonly encountered in experimental scenarios. Importantly, our proposed method eliminates the requirement for knowledge about the functional form of the distribution. Consequently, this approach offers the capability to reconstruct neutron scattering data without introducing distortions stemming from the energy-dependent cross-sections of different types of elements within materials during experimental measurements. Ultimately, this advancement facilitates a more precise interpretation and analysis of material structures based on their scattering signatures.
\end{abstract}

\clearpage

 %-------------------------------------------------------------------------
 % The main body of the paper
 %-------------------------------------------------------------------------

\section{Introduction}

Using small-angle neutron scattering (SANS) as an experimental technique to explore the intricate structural aspects of soft materials and biological systems has demonstrated remarkable success \cite{ILL}. When coupled with the contrast variation method, it has provided comprehensive insights into mesoscopic-scale structural heterogeneity, a level of detail that was previously unattainable through any other existing experimental methods to date.

The application of SANS to structural study requires the detailed knowledge of the scattering cross sections of the constituent elements within the targeted systems. More specifically, having comprehensive data for each individual isotope becomes imperative when engaging in research that employs contrast variation techniques. To thoroughly explore the relevant length scales, typical SANS experiments often use neutrons with varying wavelengths. It is well known that the total scattering cross sections exhibit distinct dependencies on the energy of neutrons employed in general SANS measurements \cite{Yip, Carpenter}. As a result, to achieve precise normalization of the scattered intensities across the entire probed reciprocal $Q$-space, it is therefore essential to account for the fluctuations in scattering cross sections resulting from variations in the incident neutron wavelength. 

SANS studies require a neutron beam with a wavelength distribution to ensure an ample neutron flux for several key reasons. Firstly, a higher flux directly enhances the strength of scattered neutrons, yielding a more detectable signal and facilitating precise data collection about the sample's structure. Secondly, a sufficient neutron flux is vital for achieving statistical accuracy, reducing measurement uncertainties, and ensuring robust experimental results. Moreover, it allows for shorter measurement times, crucial for sensitive samples. Additionally, a higher neutron flux minimizes background noise, resulting in improved signal-to-noise ratios. Ultimately, an adequate neutron flux is essential for strong signals, reduced uncertainties, efficient experiments, and minimal background noise.

The objective of this study is to develop a comprehensive mathematical framework that addresses and corrects the observed energy dependence and fluctuations in scattering cross-sections. This framework is tailored to accommodate the energy spectrum of incident neutrons relevant to general SANS studies. We establish a mathematical link by utilizing experimentally measured scattering cross-sections obtained from the National Nuclear Data Center (NNDC) at Brookhaven National Laboratory \cite{NNDC}. The specifics of this mathematical implementation are presented in the following section.

\section{Result and Discussion}\label{Sec:II}

Since a neutron has no charge it can easily enter into a nucleus and cause reaction. Neutrons interact mostly with the nucleus of an atom. As a result, one can neglect the electrons and think of atoms and nuclei interchangeably. Neutron reactions can take place at any energy, so one has to pay particular attention to the energy dependence of the interaction cross section. In the nuclear reactor neutrons with energies from $10^{-5}$ to $10^7$ eV are of interest, this means an energy range of $10^{12}$.

\begin{figure}
\centerline{
  \includegraphics[width =\columnwidth]{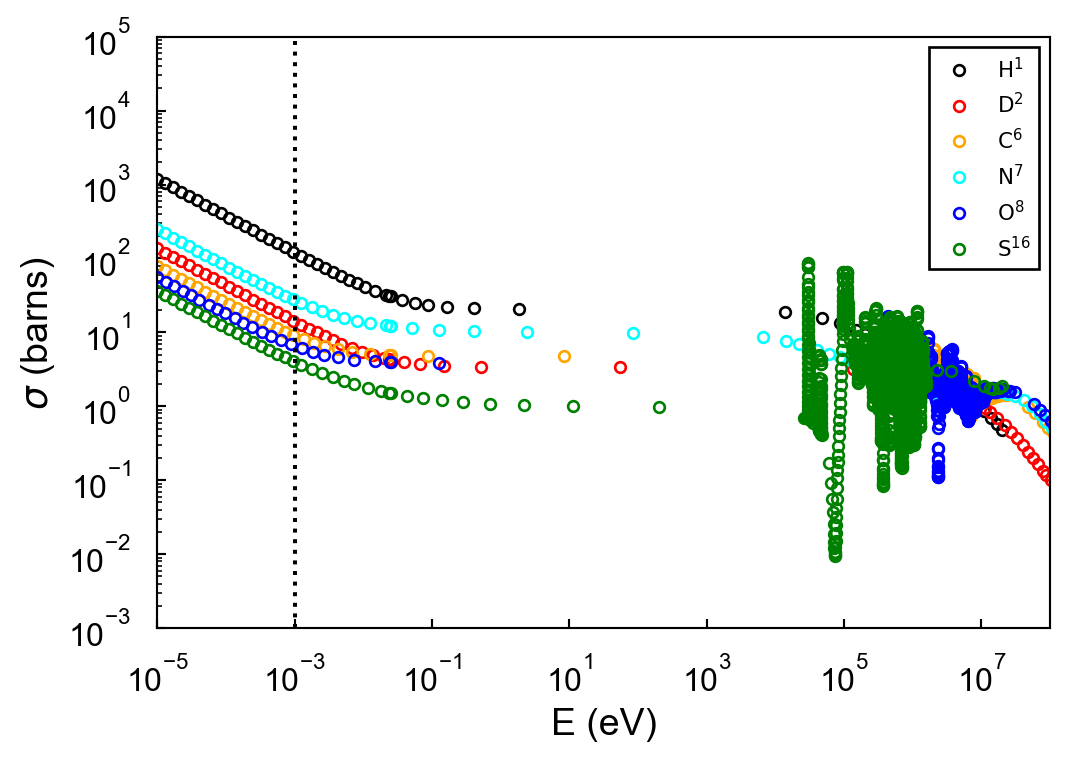}
  %\vspace{0 cm}
}  
\caption{The energy dependence of total scattering cross sections of several elements commonly encountered in soft matter and biological systems. These data have been transcribed from the National Nuclear Data Center of Brookhaven National Laboratory.
}  
\label{fig:1}
\end{figure}

Figure~\ref{fig:1} illustrates the total scattering cross sections as a function of the energy of incident neutron energy $E$, denoted as $\sigma(E)$, for various elements commonly encountered in soft materials, such as hydrogen ($\mathrm{H}^1$), deuterium ($\mathrm{D}^2$), carbon ($\mathrm{C}^6$), nitrogen ($\mathrm{N}^7$), oxygen ($\mathrm{O}^8$), and sulfur ($\mathrm{S}^{16}$). These experimental measurements were sourced from the National Nuclear Data Center at Brookhaven National Laboratory \cite{Sears, NNDC}. The superscript in the atomic notion indicates the number of protons within each nucleus.

Examining the overall characteristics of $\sigma(E)$, as illustrated in Fig.~\ref{fig:1}, yields valuable insights. To begin with, once the energy surpasses the threshold of $E > 0.1\:\mathrm{MeV}$, $\sigma(E)$ displays distinct and erratic oscillations. These observed fluctuations in scattering cross-section can be attributed to the underlying structure of compound nuclei.

\begin{figure}
\centerline{
  \includegraphics[width =\columnwidth]{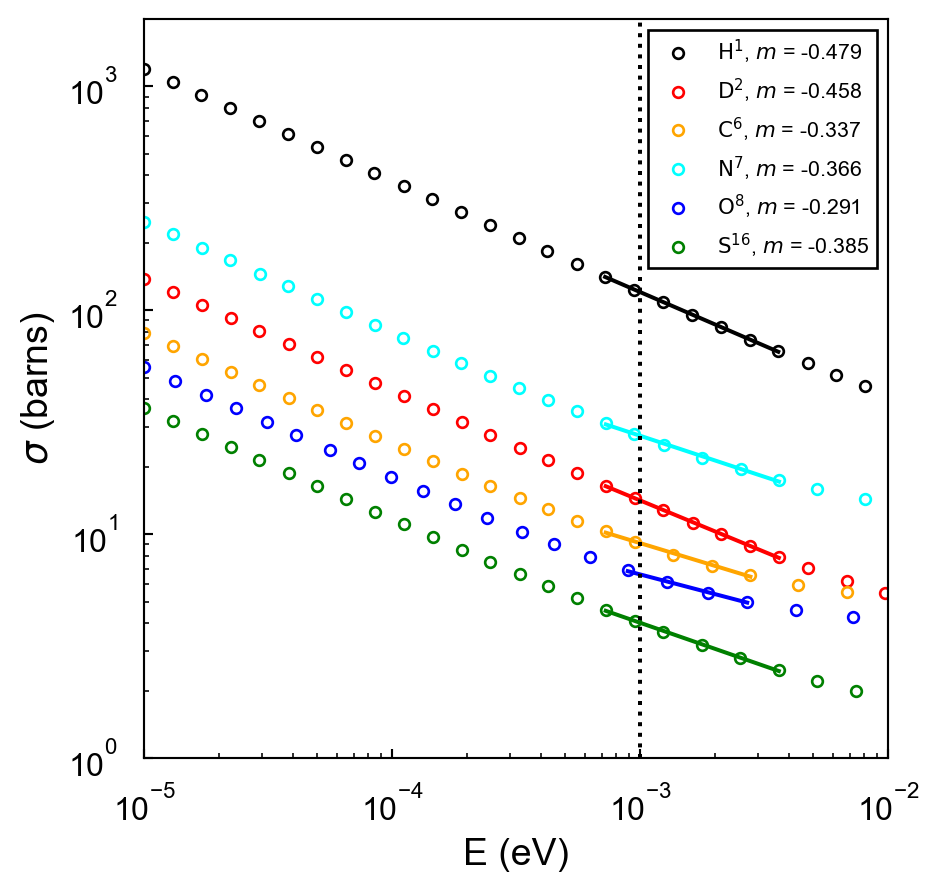}
  %\vspace{0 cm}
}  
\caption{The lower-energy region depicted in Fig.~\ref{fig:1}. In this log-log representation, the slope is seen to vary between $\sim -0.3$ to $\sim -0.5$.
}  
\label{fig:2}
\end{figure}

\begin{figure}
\centerline{
  \includegraphics[width =\columnwidth]{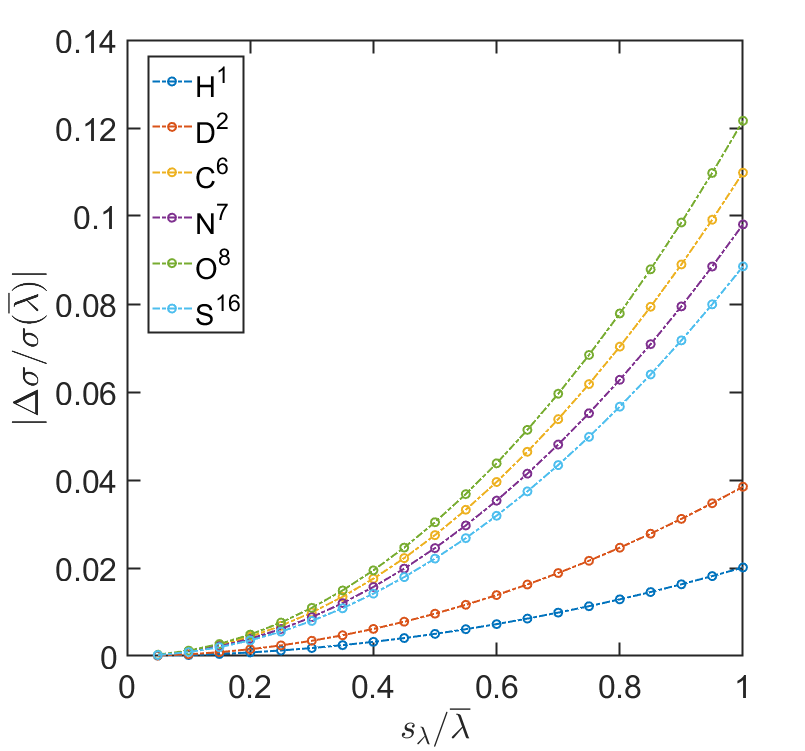}
  %\vspace{0 cm}
}  
\caption{The adjusted total scattering cross section as a function of the standard deviation of the wavelength distribution, relative to the total scattering cross sections of different elements at a specified wavelength, denoted as $\overline{\lambda}$.
}  
\label{fig:3}
\end{figure}

\begin{figure}
\centerline{
  \includegraphics[width =\columnwidth]{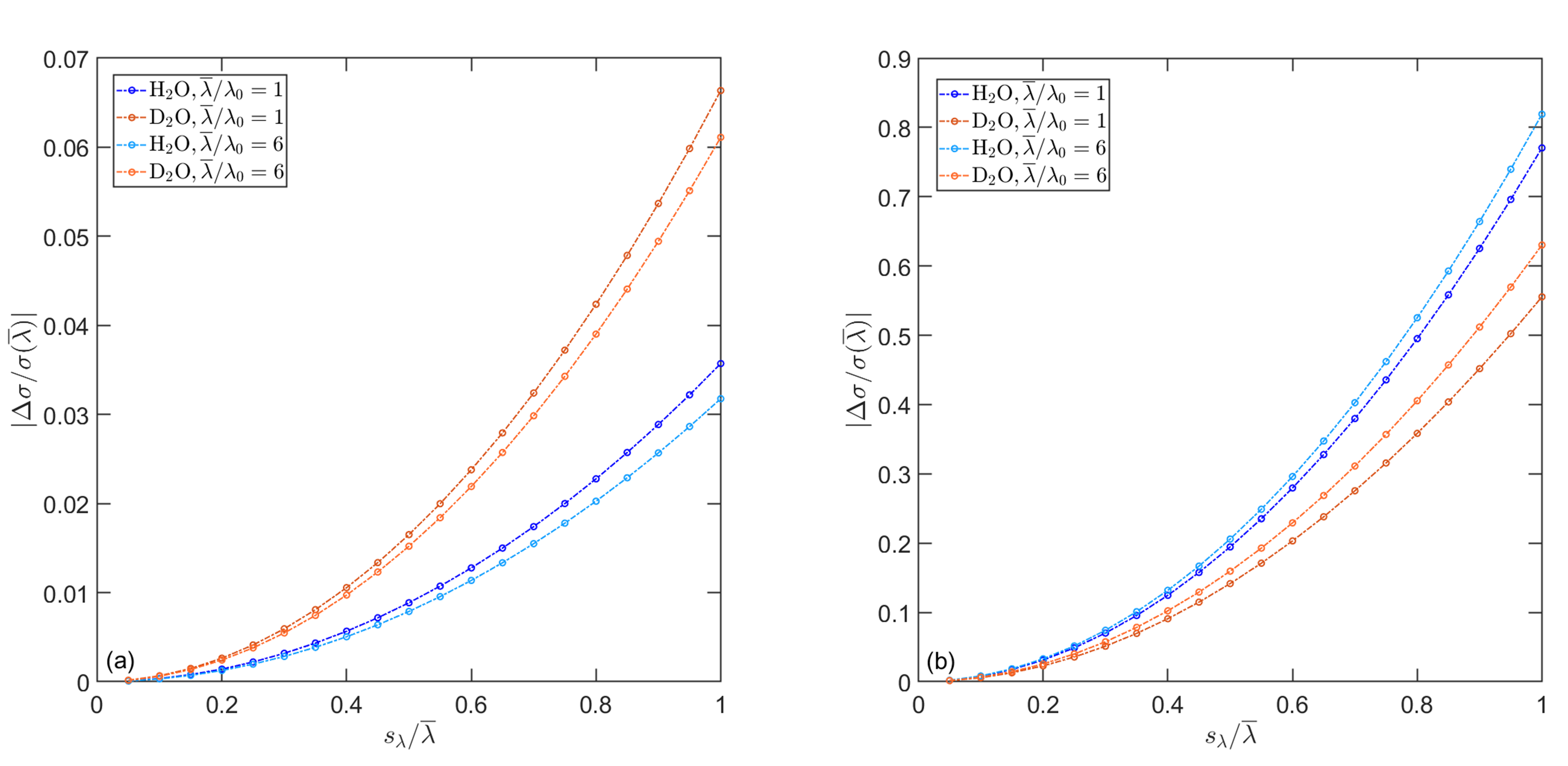}
  %\vspace{0 cm}
}  
\caption{Left panel: The adjusted total scattering cross section, considering the linear relationship on wavelength, as a function of the standard deviation of the wavelength distribution, relative to the total scattering cross sections of $\mathrm{H_2O}$ and $\mathrm{D_2O}$ at 1 $\mathrm{\AA}$ and 6 $\mathrm{\AA}$. Right panel: The adjusted total scattering cross section, accounting for the quadratic wavelength dependence, as a function of the standard deviation of the wavelength distribution, relative to the total scattering cross sections of $\mathrm{H_2O}$ and $\mathrm{D_2O}$ at 1 $\mathrm{\AA}$ and 6 $\mathrm{\AA}$.
}  
\label{fig:4}
\end{figure}

As the energy decreases, $\sigma(E)$ tends to stabilize, maintaining a relatively constant value until it reaches energies within the range of $10^4$ to $10^5$ eV. This consistent cross-section across a wide span from 0.1 MeV to 0.1 eV is commonly referred to as the free-atom cross-section. This energy-independent s-wave scattering behavior can be elucidated within the theoretical framework of the Fermi pseudopotential and the Born first approximation. Furthermore, at extremely low energy levels, there is a noticeable and consistent rise in $\sigma(E)$ as the energy ($E$) decreases from 0.1 eV onwards. As illustrated in Fig.~\ref{fig:2}, at this lower energy limit the slope is seen to vary between $-0.3$ to $-0.5$. As a result, one can establish an empirical power-law relationship between the cross-section and energy at this energy range relevant to general SANS experiments:
\begin{equation}
\ln{\frac{\sigma(E)}{\sigma_0}} = K + m\ln{\frac{E}{E_0}},
\label{eq:1}
\end{equation}
Here, $\sigma(E)$ represents the cross-section for varying neutron energies, $K$ an energy-independent dimensionless constant, and $m$ the slope when plotting $\ln{[\sigma(E)/\sigma_0]}$ against $\ln{(E/E_0)}$ at low neutron energies. $\sigma_0$ and $E_0$ serve as the normalized cross-section and energy to make the argument of the logarithmic function dimensionless.
Once $\sigma_0$ and $E_0$ are defined, these variables can be extracted via regression analysis using experimental data for different elements. Figure~\ref{fig:2} presents the log-log plot of $\sigma(E)$ versus $E$ for hydrogen, deuterium, carbon, nitrogen, oxygen, and sulfur. Overall, each element presented here has the similar trend, and the values of $m$ are roughly the same. Thus, the power-law relationship can be expressed as:
\begin{equation}
\sigma(E) = A\sigma_0 \left(\frac{E}{E_0}\right)^m,
\label{eq:2}
\end{equation}
where $A$ is defined as $A=\exp{(K)}$. Typically, $\sigma_0 = 10^{-24}\: \mathrm{cm}^2$ and $E_0 = 1\:\mathrm{eV}$, as depicted in Fig.~\ref{fig:1}.
Through the following Einstein-de Broglie relationship \cite{Schiff},
\begin{equation}
E = \frac{p^2}{2m_n} = \frac{h^2}{2m_n\lambda^2} =  \frac{h^2c^2}{2m_nc^2\lambda^2} \approx 0.0818 (\lambda_0/\lambda)^2 \: (\mathrm{eV}), 
\label{eq:3}
\end{equation}
where $h$ is the Planck's constant, $c$ the speed of light, $m_n$ the rest mass of a neutron, and $\lambda_0 = 1\:\mathrm{\AA}$, we can convert the energy dependence with $E_0 = 1 \:\mathrm{eV}$ into wavelength dependence:
\begin{equation}
\sigma(\lambda) = 0.0818^m A \sigma_0 \left(\frac{\lambda_0}{\lambda}\right)^{2m},
\label{eq:4}
\end{equation}
Since in both reactor-based and time-of-flight neutron sources, the wavelength distributions of incident neutrons typically do not follow a Dirac-delta function, it's essential to consider the wavelength effect when calculating the cross-section. For a given distribution $f(\lambda)$, the summation of contributions from different wavelengths can be calculated straightforwardly through integration:
\begin{equation}
\sigma_c = \int d\lambda f(\lambda) \sigma(\lambda).
\label{eq:5}
\end{equation}
In general neutron experiments, the wavelength distributions are characterized by the mean and standard deviation of wavelength, defined as follows:
\begin{equation}
\overline{\lambda} = \int d\lambda f(\lambda) \lambda, \quad s_{\lambda}^2 = \int d\lambda f(\lambda) (\lambda-\overline{\lambda})^2
\label{eq:6}.
\end{equation}
Hence, as long as $s_{\lambda}$ remains sufficiently small, we can approximate Eqn.~\eqref{eq:5} using the central moment expansion as:
\begin{equation}
\sigma_c \approx \int d\lambda f(\lambda)[\sigma(\overline{\lambda}) + \sigma'(\overline{\lambda})(\lambda-\overline{\lambda}) + \frac{\sigma''(\overline{\lambda})}{2}(\lambda-\overline{\lambda})^2] = \sigma(\overline{\lambda}) + \frac{\sigma''(\overline{\lambda})}{2} s_{\lambda}^2.
\label{eq:7}
\end{equation}
This approximation provides a practical method for computing cross-sections in neutron experiments that involve non-negligible wavelength distributions. From Eqn.~\eqref{eq:7}, we observe that the correction is primarily driven by the variance of the wavelength distribution:
\begin{equation}
    \Delta \sigma = \sigma_c - \sigma(\overline{\lambda}) \approx \frac{\sigma''(\overline{\lambda})}{2} s_{\lambda}^2
    = m(2m+1)\sigma(\overline{\lambda}) (\frac{s_\lambda}{\overline{\lambda}})^2.
    \label{eq:8}
\end{equation}
From Eqn.~\eqref{eq:8}, it is evident that the correction is proportional to the square of $s_\lambda/\overline{\lambda}$ and depends on the exponent $m$. Therefore, this approximation holds valid when $\overline{\lambda}$ significantly exceeds $s_{\lambda}$, which is typically the case in both reactor-based and spallation neutron sources. However, spallation sources generally exhibit a smaller ratio of $s_{\lambda}/\overline{\lambda}$ compared to reactor-based sources. Since the exponent $m$ for each element is inherently of a similar magnitude, as indicated by the slope in the log-log plot in Fig.~\ref{fig:2}, the correction is primarily determined by the variance of the wavelength distribution. Consequently, the relative correction concerning $\sigma(\overline{\lambda})$ can be straightforwardly calculated as:
 \begin{equation}
     |\frac{\Delta\sigma}{\sigma(\overline{\lambda})}| = |m(2m+1)|(\frac{s_\lambda}{\overline{\lambda}})^2
     \label{eq:9}
 \end{equation}
In Fig.~\ref{fig:3}, we illustrate the relative correction as a function of $s_{\lambda}/\overline{\lambda}$ for the elements shown in Fig.~\ref{fig:2}. Due to the nature of the power-law decay, this contribution becomes substantial, exceeding $3\:\%$ and $12\:\%$ respectively for deuterium and oxygen, when $s_{\lambda}/\overline{\lambda}$ approaches unity. Once more, it's important to note that this scenario is not characteristic of typical reactor-based and spallation neutron sources. The validity of this approximation remains practical and relevant for neutron scattering measurements. In the case of a molecule, its scattering behavior is the cumulative effect of all the scattering lengths of the atoms it contains. Therefore, we can calculate its scattering length, denoted as `$b$', as follows:
\begin{equation}
    b = \sum_i b_i = \sqrt{\frac{1}{4\pi}}\sum_i \sqrt{\sigma_i}
    \label{eq:10}
\end{equation}
The corresponding cross-section for a molecule can be expressed as:
\begin{equation}
    \sigma = 4\pi b^2 = \sum_{i} \sigma_{c,i} + \sum_{i,j} (1-\delta_{ij}) \sqrt{\sigma_{c,i}\sigma_{c,j}},
    \label{eq:11}
\end{equation}
Here, $\sigma_{c,i}$ and $\sigma_{c,j}$ represent the corrected cross-sections for $i-$ and $j-$th atoms, respectively, and their values can be directly computed using Eqn.~\eqref{eq:7}. Note that the cross term in Eqn.~\eqref{eq:11} needs to be integrated after first calculating the square root of the cross section with respect to the wavelength distribution.
One of the most paramount examples in neutron scattering experiments is the case of normal and heavy water, $\mathrm{H_2O}$ and $\mathrm{D_2O}$, respectively.  The corrected cross-sections for these molecules can be calculated using Eqn.~\eqref{eq:7} and Eqn.~\eqref{eq:11}, as shown below:
\begin{equation}
    \sigma_{c,\mathrm{H_2O}} = 4\sigma_{c,\mathrm{H}} + \sigma_{c,\mathrm{O}} + 4 \sqrt{\sigma_{c,\mathrm{H}}\sigma_{c,\mathrm{O}}},
     \sigma_{c,\mathrm{D_2O}} = 4\sigma_{c,\mathrm{D}} + \sigma_{c,\mathrm{O}} + 4 \sqrt{\sigma_{c,\mathrm{D}}\sigma_{c,\mathrm{O}}}
     \label{eq:12}
\end{equation}
These relative corrected results are presented in Fig.~\ref{fig:4} as a function of both $\overline{\lambda}$ and $s_{\lambda}/\overline{\lambda}$. In a  general SANS measurement, various wavelength-dependent factors influencing the recorded scattering cross-sections have been identified \cite{Frielinghaus, Carpenter}. These factors include elastic incoherent scattering, which exhibits no wavelength dependency, inelastic incoherent scattering, which is directly proportional to wavelength, and coherent elastic SANS scattering, which is proportional to the square of the wavelength. It is also essential to note that different configurations and variances of neutron wavelength distributions at neutron beamlines can lead to varying cross-section magnitudes. When conducting structural analysis of neutron scattering data, it is imperative to account for these factors to ensure an unbiased interpretation.

The relative correction of $\frac{\Delta \sigma}{\sigma (\overline{\lambda})}$, which takes into account the dependence of $\lambda^{2m}$ on the cross-section in Fig.~\ref{fig:2}, and its relationship with the wavelength distribution for $\mathrm{H_2O}$ and $\mathrm{D_2O}$ are shown in the left panel of Fig.~\ref{fig:4}. For both cases with $-2m$ roughly on the magnitude of $1$, the mean wavelength is selected as 1 $\mathrm{\AA}$ and 6 $\mathrm{\AA}$, respectively. Firstly, the cross section of $\mathrm{H_2O}$ is greater than that of $\mathrm{D_2O}$, primarily because the cross section of $\mathrm{H}$ is larger than that of $\mathrm{D}$, as illustrated in Fig.~\ref{fig:2}. Hence, the relative correction of $\mathrm{H_2O}$ would be small compared to that of $\mathrm{D_2O}$. Secondly, as expected, the significance of the correction increases as the wavelength becomes longer and the wavelength distribution becomes widens. The right panel displays the corrected $\frac{\Delta \sigma}{\sigma \overline{\lambda}}$ for $\mathrm{H_2O}$ and $\mathrm{D_2O}$, which considers the dependence of $\lambda^{-4m}$ on the cross-section with $-4m$ roughly on the magnitude of $2$, and illustrates its connection with the wavelength distribution. We observe a qualitative trend similar to the results shown in the left panel. However, quantitatively, the magnitude of the correction is an order of magnitude larger than the correction presented in the left panel. In this low-energy limit, where the incident neutron energy is on par with the thermal energy of the target nuclei, the assumption that the target nuclei are stationary is no longer valid. To consider the thermal movements of the target, it becomes essential to understand the physical state of the target, such as a crystalline or a liquid target. Our findings presented in Fig.~\ref{fig:4} indicate the critical importance of incorporating the dynamics of the target nuclei into the calculation of total scattering cross sections. 

\section{Conclusion}\label{Sec:III}

A thorough grasp of the energy dependence of neutrons serves as a foundational cornerstone when it comes to the successful execution, meaningful interpretation, and precise analysis of Small-Angle Neutron Scattering (SANS) experiments. This profound understanding ensures that the data generated through these experiments not only meet the standards of accuracy but also unlock valuable insights into the intricate structure and unique properties of the materials under investigation.

Within the context of this report, we have dedicated ourselves to the development of a systematic and rigorous methodology. This methodology is specifically designed to rectify and enhance the accuracy of scattering cross-sections derived from experimental SANS data. By addressing the complexities associated with neutron energy dependence head-on, we are advancing the field of SANS research, ultimately empowering scientists and researchers to extract richer, more detailed information from their experiments. This newfound precision opens doors to deeper comprehension of the materials being studied, paving the way for breakthroughs and innovations in various scientific domains where SANS plays a pivotal role.

Our proposed approach for energy correction not only fulfills the critical need for precise data interpretation, especially in response to the energy-dependent behaviors observed in SANS detectors but also offers broader applicability.

Furthermore, this approach can be extended to tackle the complexities introduced by multiple scattering effects in SANS experiments. By comprehensively accounting for neutron energy dependence, our method aids in accurately quantifying and correcting these multiple scattering contributions, enhancing the fidelity of SANS measurements.

Additionally, the versatility of our approach extends to SANS instrument calibration. In this context, where precise calibration is essential, a deep understanding of how neutron energy influences instrument response is paramount. By incorporating neutron energy considerations, our method contributes to the refinement of instrument calibration procedures, ensuring the accuracy and reliability of SANS instruments across a wide range of experimental conditions.

\clearpage

     % Appendices appear after the main body of the text. They are prefixed by
     % a single \appendix declaration, and are then structured just like the
     % body text.

%\appendix
%\section{Appendix title}

%\subsection{Title}

%\subsubsection{Title}

     %-------------------------------------------------------------------------
     % The back matter of the paper - acknowledgements and references
     %-------------------------------------------------------------------------

     % Acknowledgements come after the appendices

\ack{Acknowledgements}

This research was performed at the Spallation Neutron Source, which is a DOE Office of Science User Facility operated by Oak Ridge National Laboratory. KEA extend her heartfelt appreciation to the ORNL Next Generation Science, Technology, Engineering, and Mathematics (STEM) Internship Program (NGSI) for their financial support during my summer 2023 internship at SNS, ORNL.

\clearpage

\bibliographystyle{iucr}
\bibliography{references}

     %-------------------------------------------------------------------------
     % TABLES AND FIGURES SHOULD BE INSERTED AFTER THE MAIN BODY OF THE TEXT
     %-------------------------------------------------------------------------

%---------------------------------------------
\clearpage
%---------------------------------------------
%---------------------------------------------
%---------------------------------------------
%---------------------------------------------
%---------------------------------------------
%---------------------------------------------
%---------------------------------------------
\clearpage

\end{document}